# LEVEL SHIFTER DESIGN FOR LOW POWER APPLICATIONS

Manoj Kumar[1], Sandeep K. Arya[1], Sujata Pandey[2]

[1]Department of Electronics & Communication Engineering
Guru Jambheshwar University of Science & Technology, Hisar, India,
manojtaleja@rediffmail.com, arya1sandeep@rediffmail.com
[2]Amity University, Noida, India
spandey@amity.edu

## ABSTRACT

*With scaling of Vt sub-threshold leakage power is increasing and expected to become significant part of total power consumption.In present work three new configurations of level shifters for low power application in 0.35µm technology have been presented. The proposed circuits utilize the merits of stacking technique with smaller leakage current and reduction in leakage power. Conventional level shifter has been improved by addition of three NMOS transistors, which shows total power consumption of 402.2264pW as compared to 0.49833nW with existing circuit. Single supply level shifter has been modified with addition of two NMOS transistors that gives total power consumption of 108.641pW as compared to 31.06nW. Another circuit, contention mitigated level shifter (CMLS) with three additional transistors shows total power consumption of 396.75pW as compared to 0.4937354nW. Three proposed circuit's shows better performance in terms of power consumption with a little conciliation in delay. Output level of 3.3V has been obtained with input pulse of 1.6V for all proposed circuits.*

## 1. INTRODUCTION

With the growing demand of handheld devices like cellular phones, multimedia devices, personal note books etc., low power consumption has become major design consideration for VLSI circuits and system [1], [2]. With increase in power consumption, reliability problem also rises and cost of packaging goes high [3]. Power consumption in VLSI circuit consists of dynamic and static power consumption. Dynamic power has two components i.e. switching power due to the charging and discharging of the load capacitance and the short circuit power due to the non-zero rise and fall time of the input waveforms [4]. The static power of CMOS circuits is determined by the leakage current through each transistor. Power consumption of VLSI circuits can be reduced by scaling supply voltage and capacitance [4]. With the reduction in supply voltage, problems of small voltage swing, insufficient noise margin and leakage currents originate [5]. With the development of technology towards submicron region leakage power has become significant component of total power dissipation [6], [7]. Static power component of power consumption must be given due consideration if current trends of scaling of size and supply voltage need to be sustained.

In System on chip (SoC) design, different parts like digital, analog, passive component are fabricated on a single chip and needs different voltages to achieve optimum performance. Level converters are used to convert the logic signal from one voltage level to other level and are the significant circuit component in VLSI systems. Level shifters are also important circuit component in multi voltage systems and have been used in between core circuits and I/O circuit. Various design for level shifters have been reported in literature with single and dual supply [8]- [16]. Conventional level shifter using 10 transistor with low voltage supply VddL and high voltage supply VddH has been reported [8], [10], [11], [12]. The conventional level shifters have disadvantages of delay variation due to different current driving capabilities of transistors, large power consumption and failure at low supply core voltage VddL [11]. The single supply level shifter allows communication between modules without adding any extra supply pin. Single supply level shifters have advantages over dual supply in terms of pin count, congestion in routing and overall cost of the system. Another benefit of single supply is flexible placement and routing in physical design. Single supply level shifters dissipate higher leakage power due to increase in leakage currents when input supply level is lower or VddH is higher than input supply level by more than $V_{tn}$ [12]. Contention mitigated level shifter (CMLS) using 12 transistors with reduced power consumption and delay than conventional level shifter has been reported [13]. Conventional level converters using bootstrapped gate drive to reduce voltage swings and power consumption has been reported [8]. In [14] method to modify the threshold voltage for reduce power consumption using dual supply voltage has been reported.

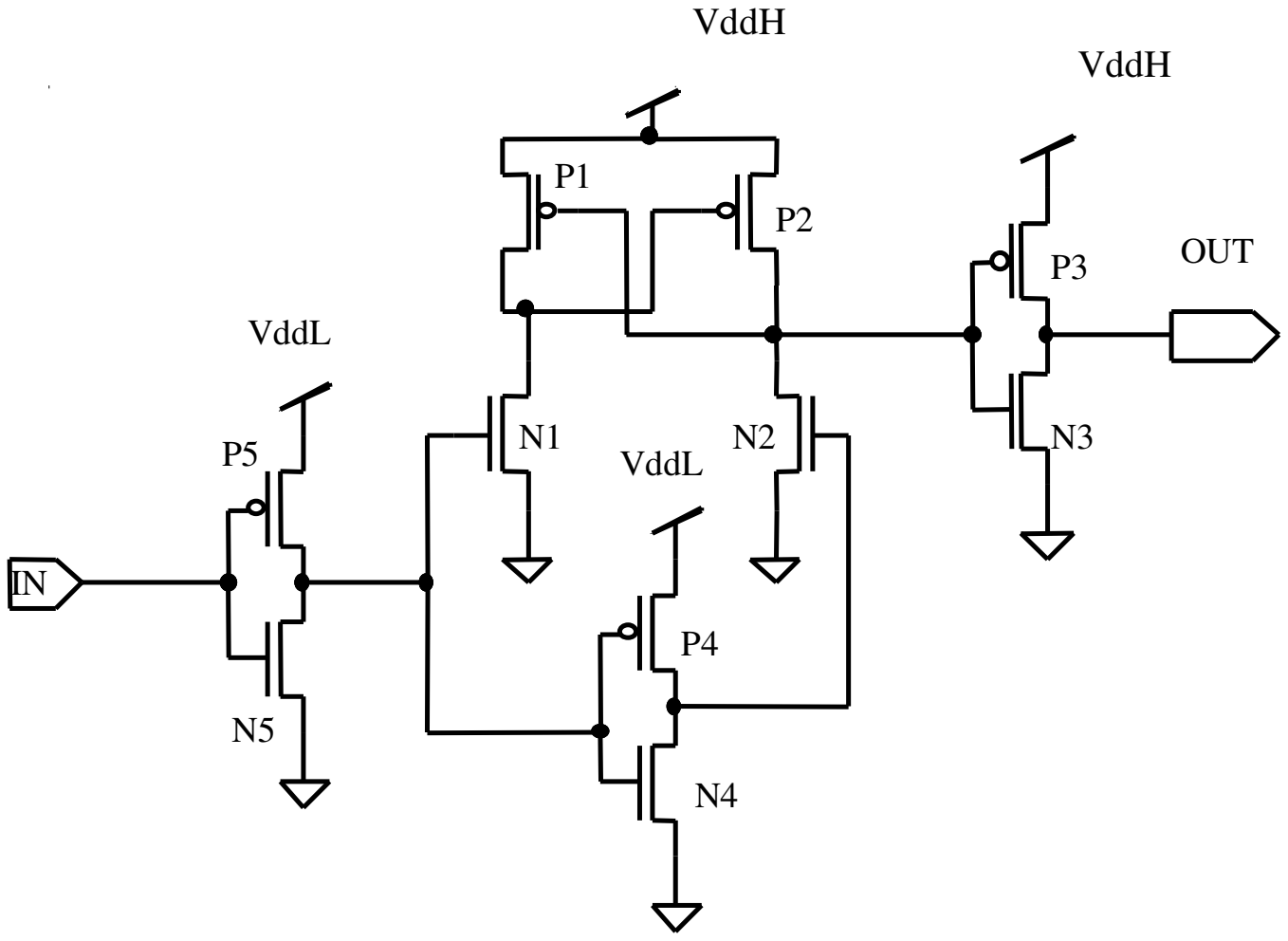

(a)

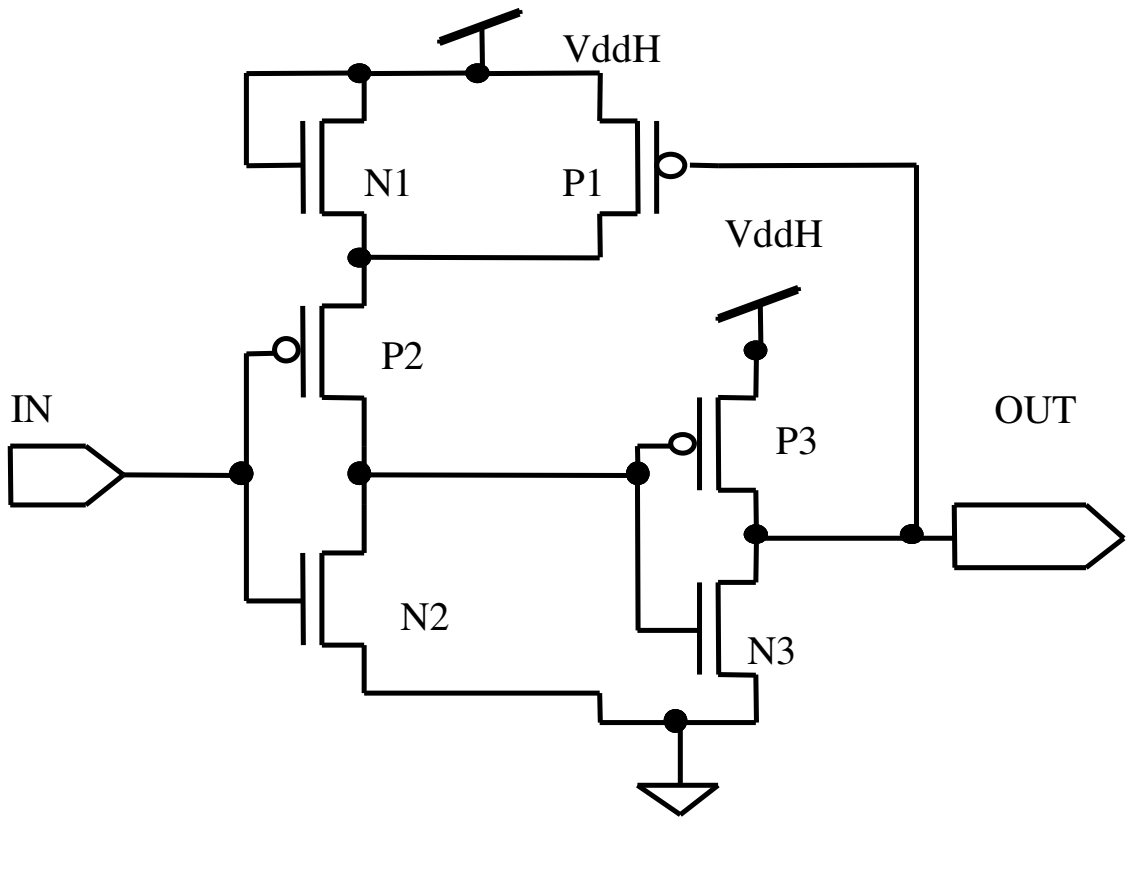

(b)

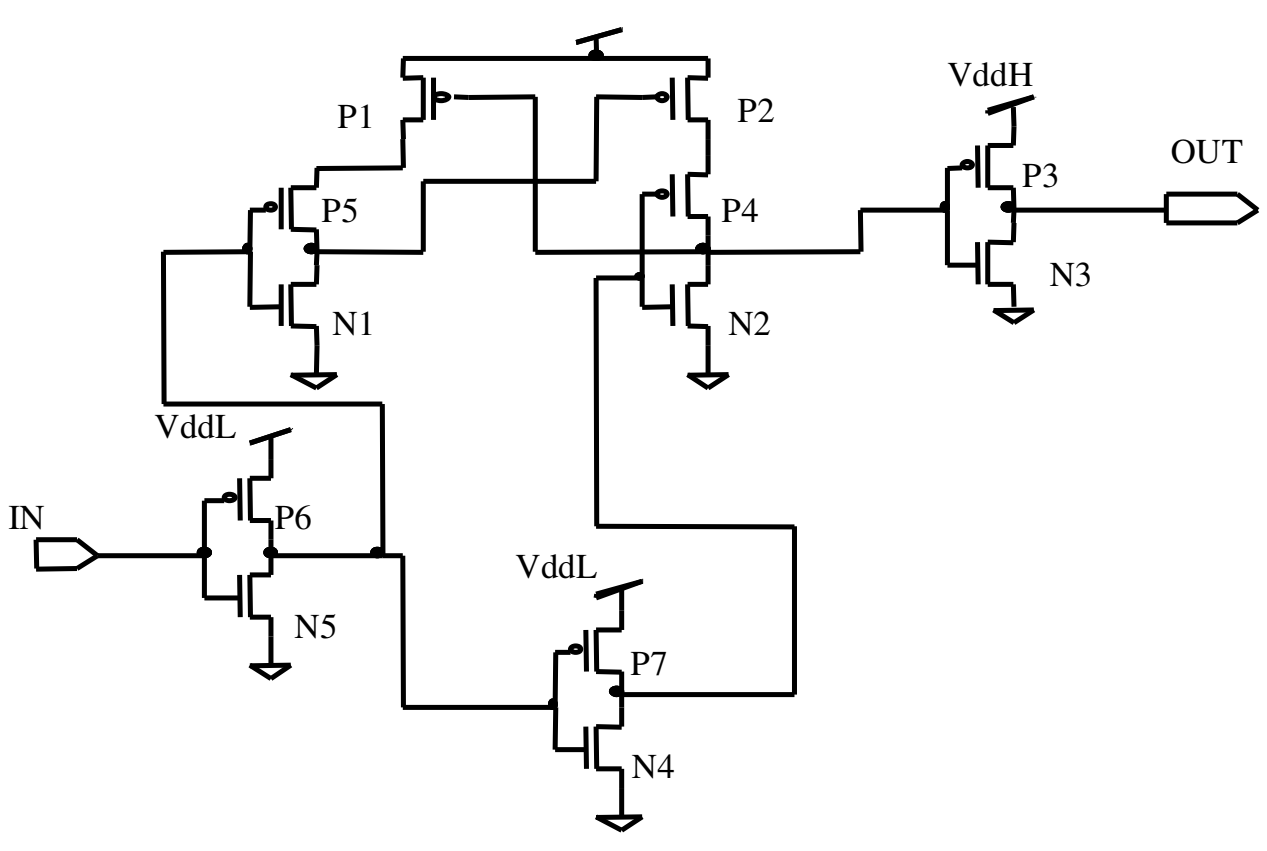

(c)

Figure.1 Level shifter circuits (a) Conventional (b) Single supply (c) Contention mitigated

With increase in operating frequency and number of level shifters in data driver's circuits, power consumption has become major performance metrics. It has been reported that stacking of two off devices reduces the sub-threshold leakage as compared to single off device [7], [17]-[20]. In the current work an effort has been made to reduce the leakage power consumption of level shifter circuits using the concept of stacking technique without compromising the outputs levels. Rest of the paper is organized as follows: in Section II stacking technique has been applied to existing circuits and modified circuits have been presented. In Section III the results of modified circuits have been compared with earlier existing circuits. Conclusions have been drawn in Section IV.

## 2. SYSTEM DESCRIPTION

In present work, modifications have been proposed in existing level converter circuits namely conventional, single supply, and contention mitigated for improvement in power dissipation. Conventional level shifter with stacking uses three additional NMOS transistors as shown in Fig. 2. Three NMOS transistors [N3-N5] of conventional level shifter with gate length 0.35µm and width 1.0µm has been replaced by six transistors [N3-N8] with same gate length and width of 0.5µm. Gate lengths of all NMOS and PMOS transistors have been taken as 0.35µm. Normal values of widths 1.0 and 2.5µm for NMOS transistors [N1&N2] and PMOS transistors [P1-P5] have been taken. Supply voltage VddH and VddL are taken as 3.3 V & 2.2V respectively.

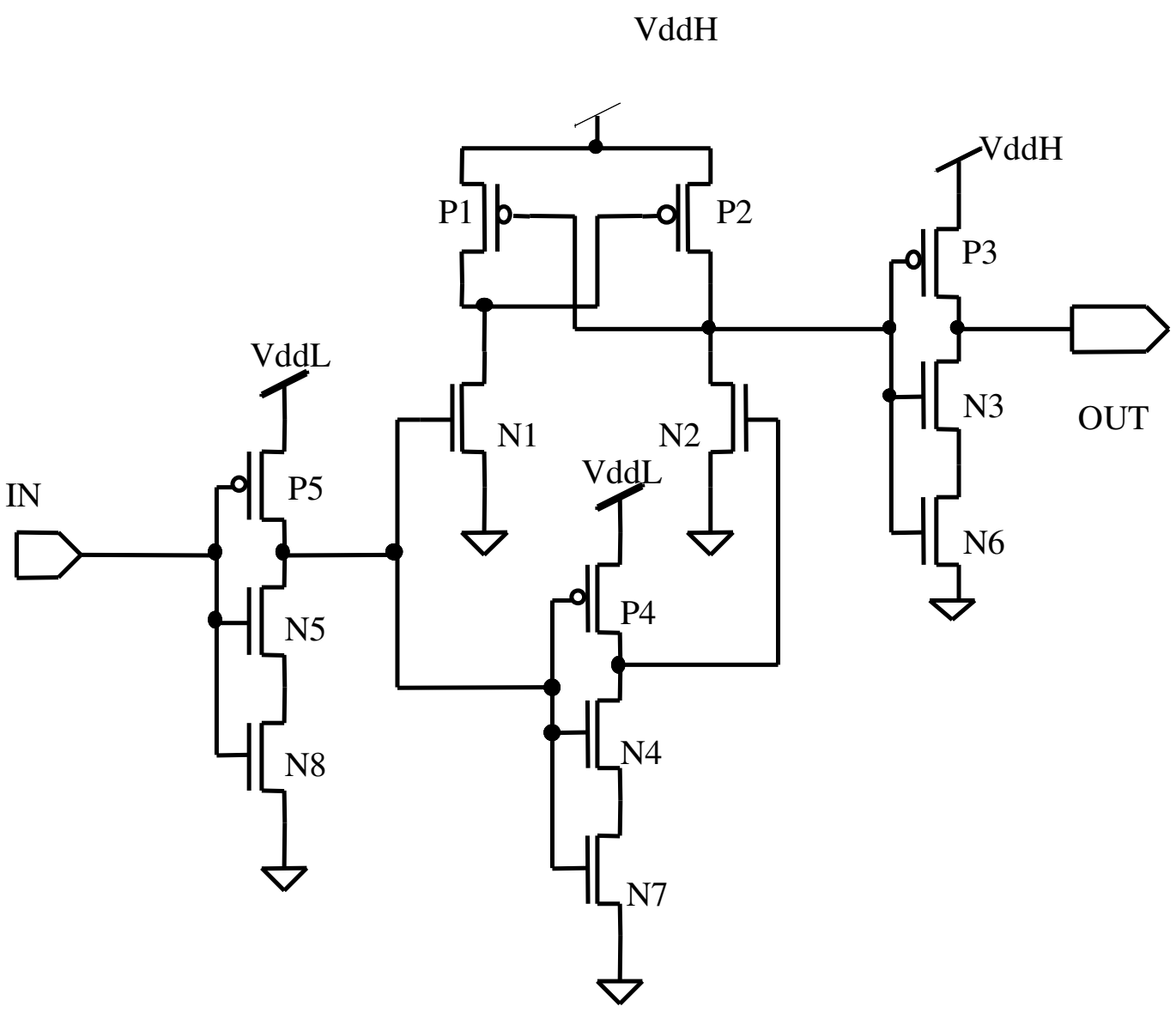

Figure.2 Conventional level shifter with stacking technique

Fig.3 shows modified single supply level shifter with stacking technique using two additional NMOS [N4-N5] transistors. NMOS transistors [N2-N3] with gate length 0.35µm and width 1.0µm have been replaced by four transistors [N2-N5] same gate length and width of 0.5µm. Gate lengths of all transistors have been taken as 0.35µm. Width ($W_n$) for [N1-N5] has been taken as 0.5µm, preserving total width 1.0 µm. Normal values of widths 2.5µm have been taken for PMOS [P1-P3]. Supply voltage VddH has been taken as 3.3 V.

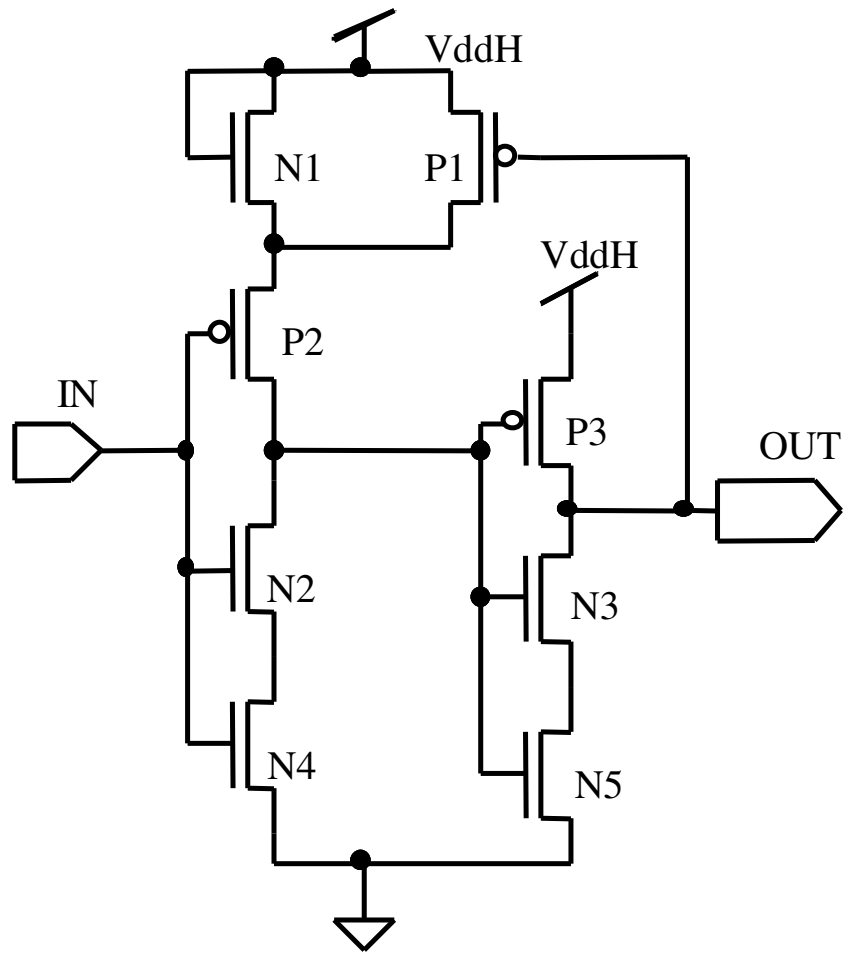

Figure.3 Single supply level shifter with stacking technique

Fig.4 shows modified contention mitigated level shifter employing stack forcing with addition of three NMOS transistors [N6-N8]. NMOS transistors [N3-N5] with gate length 0.35µm and width 1.0µm have been replaced by six transistors [N3-N8] with same gate length and width of 0.5µm preserving the total width 1.0 µm. Gate lengths of all transistors have been taken as 0.35 µm. Normal values of widths 1.0 and 2.5µm have been taken for NMOS [N1&N2] and PMOS [P1-P7] transistors respectively. Supply voltages VddH and VddL have been taken as 3.3V and 2.2V respectively. Level shifter circuits shown in Fig.1 also have been designed with gate lengths of 0.35μm and widths of PMOS & NMOS have been taken as 2.5μm & 1.0μm respectively.

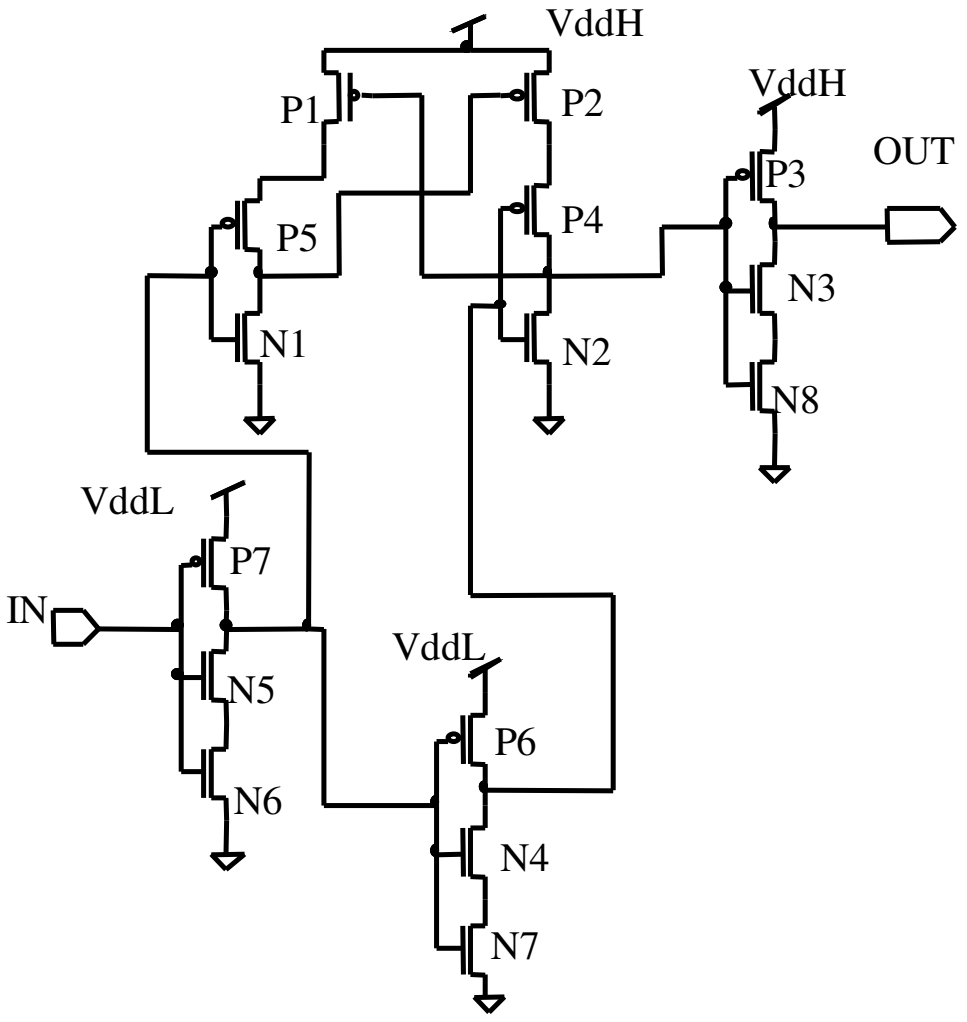

Figure.4 Contention mitigated level shifter with stacking technique

# 3. RESULTS AND DISCUSSIONS

Modified level shifter circuits [Fig.2-4] with stack forcing have been presented and simulated in 0.35μm technology using TSMC0.35 model file. Table I shows the results for existing level shifter and Table II shows results of modified circuits. Modified conventional level shifter gives power consumption of 402.2264pW as compared to 0.49833nW with existing conventional circuit. Modified single supply level shifter shows 108.641pW compared to 31.06nW with existing circuit. Finally, the modified CMLS shows 396.75pW as compared to 0.4937354nW without modifications. Results show that power consumption has been reduced in modified circuits with application of stacking technique. Delays of existing and proposed circuits also have been obtained and shown in Table I&II. Fig.6 (a) and (b) shows power consumptions and delay of proposed level shifters circuits. For comparisons existing circuits have been simulated with same set of parameters as for proposed circuits. Fig.7 (a) and (b) shows power consumptions and delay of existing level shifters circuits. Results show that three proposed circuit's shows better performance in terms of power consumption with a little conciliation in delay.

Table-I Results for proposed circuits

| **Level shifter configurations** | **Power Consumption (pW)** | **Delay (ns)** |
|---|---|---|
| Modified conventional level shifter | 402.2264 | 2.3376 |
| Modified single supply level shifter | 108.641 | 2.564 |
| Modified contention mitigated level shifter | 396.75 | 0.55206 |

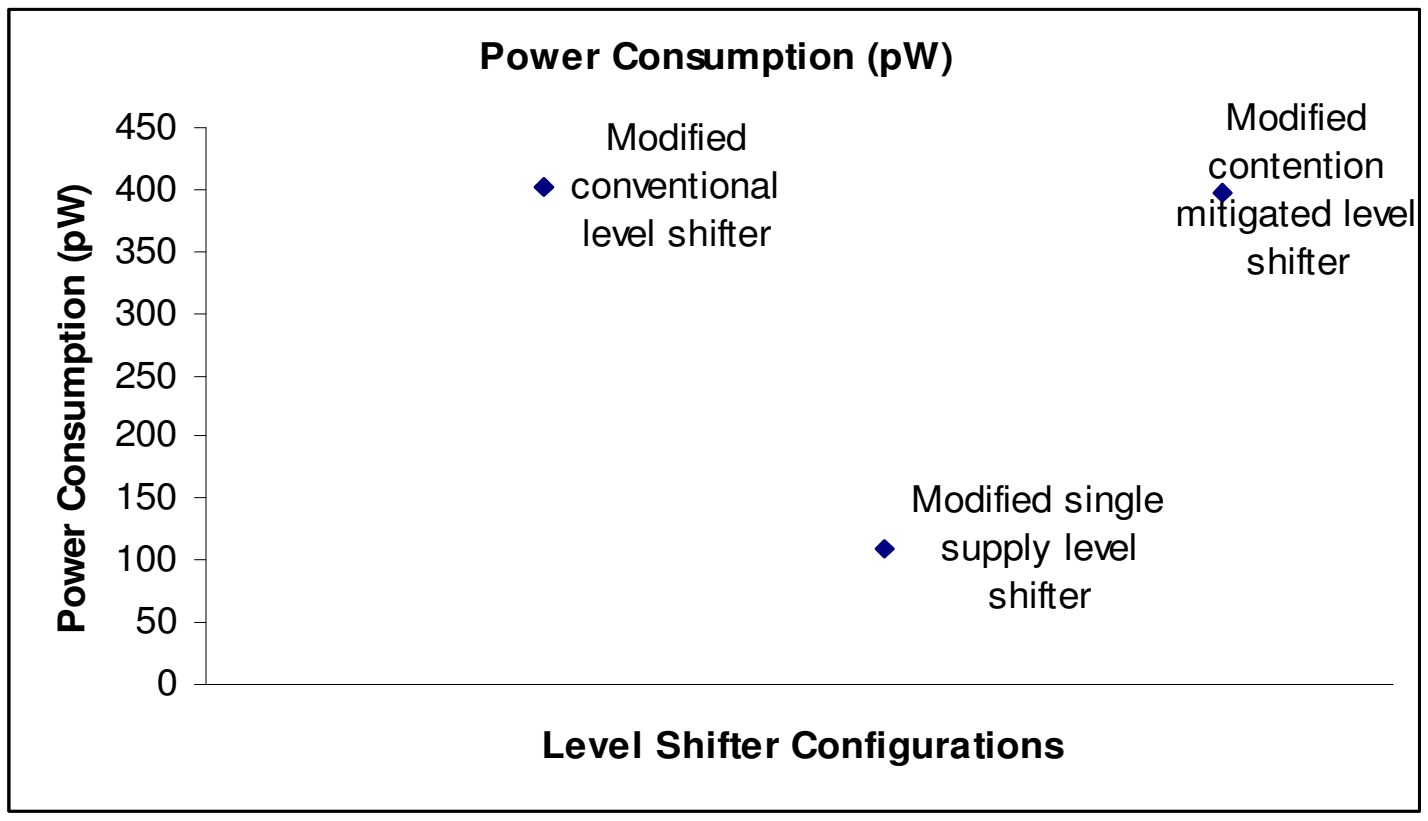

(a)

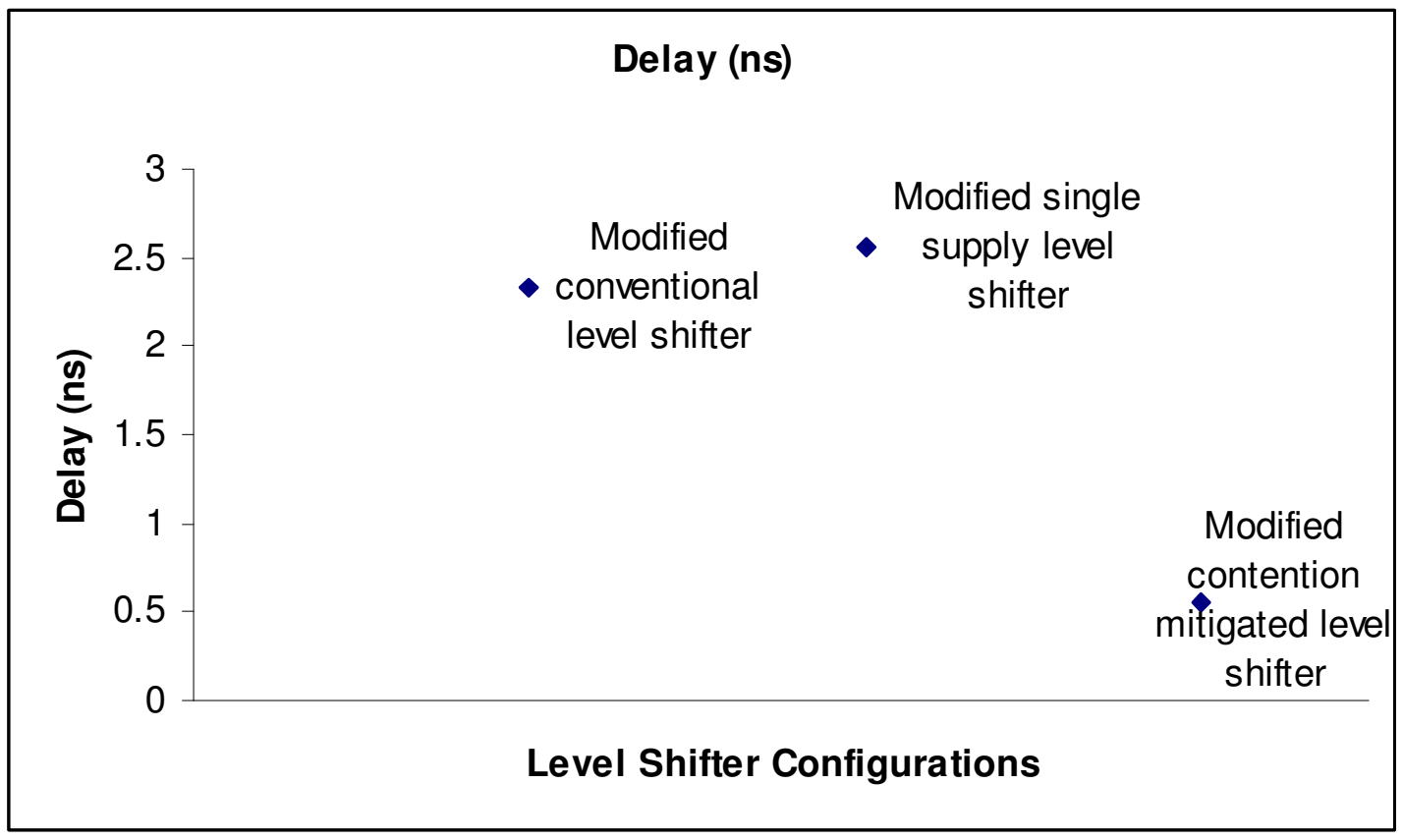

(b)

Figure.6 (a) power consumption (b) delay of proposed circuits

Table-II Results for existing level shifters

| **Level shifter configurations** | **Power Consumption (nW)** | **Delay(ns)** |
|---|---|---|
| Conventional level shifter[11] | 0.49833 | 2.2744 |
| Single supply level shifter[12] | 31.06 | 0.33474 |
| Contention mitigated level shifter[13] | 0.4937354 | 0.391815 |

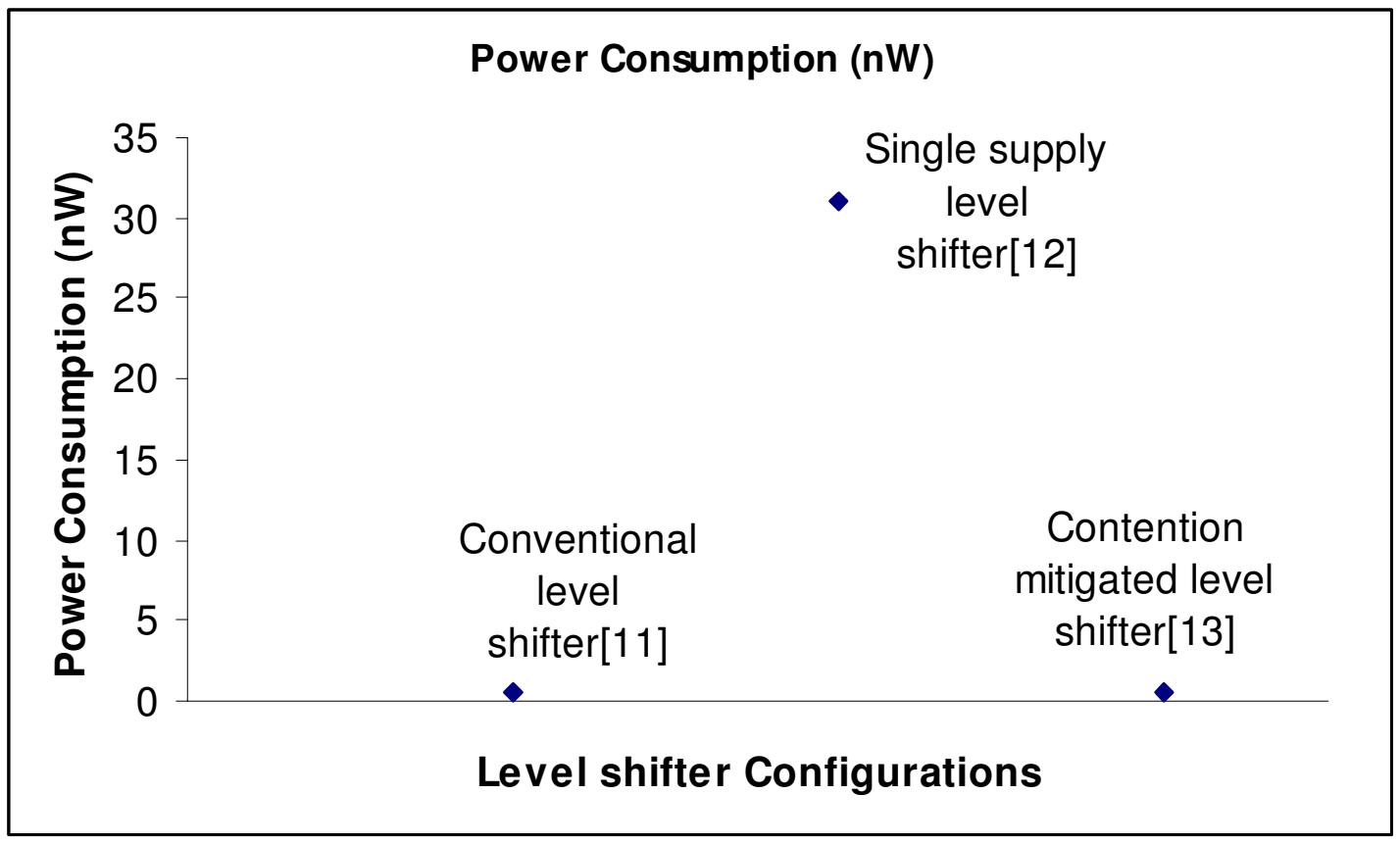

(a)

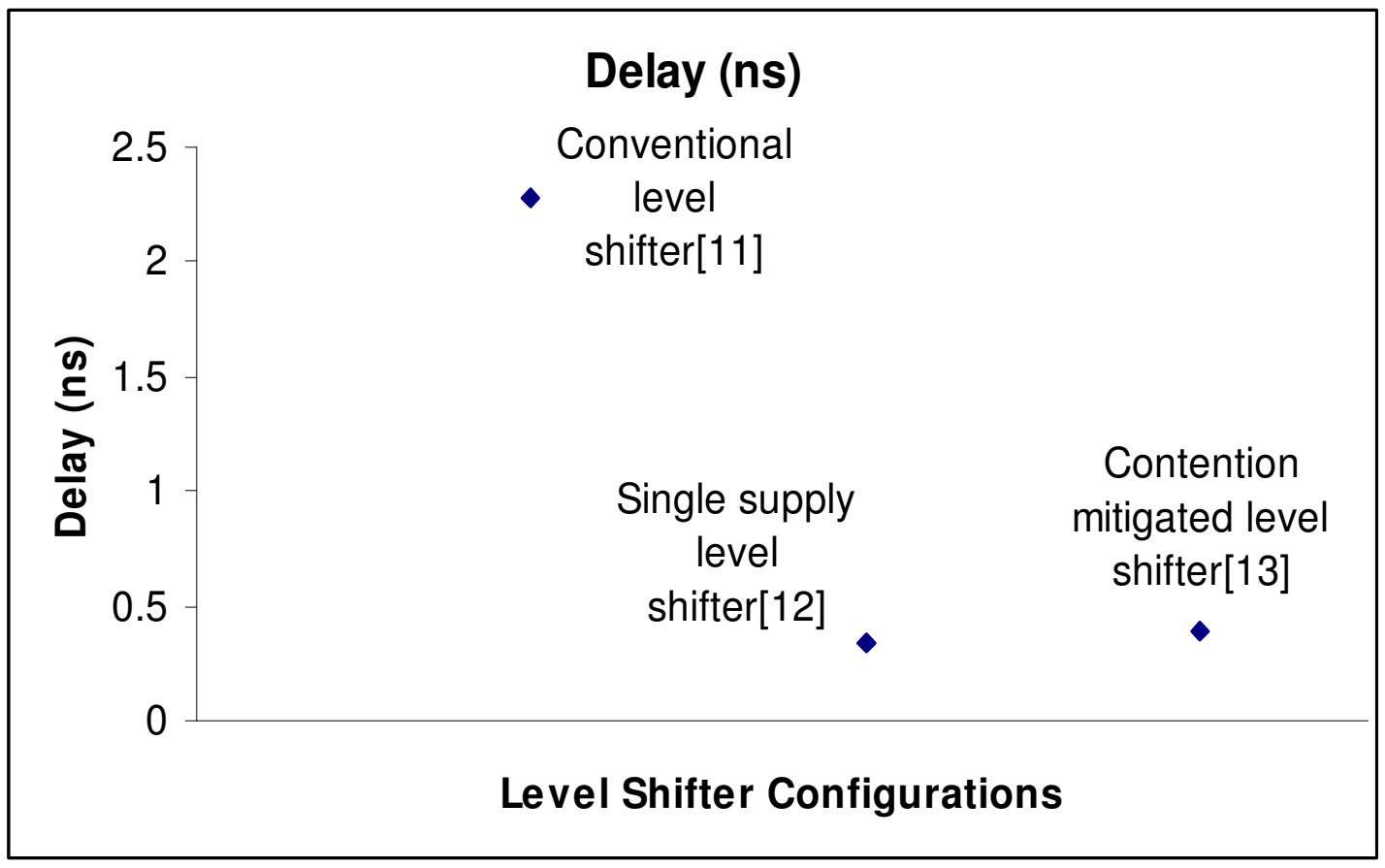

(b)

Figure.7 (a) power consumption (b) delay of existing circuits

## 4. CONCLUSIONS

In present paper three new circuits of level shifters namely modified conventional, modified single supply and modified contention mitigated have been presented. Modified conventional level shifter gives power consumption of 402.2264pW as compared to 0.49833nW for conventional level shifter. Proposed single supply shows power consumption of 108.641pW as compared to 31.06nW for conventional single supply. Third proposed circuit's shows power consumption of 396.75pW as compared to 0.4937354nW for existing circuit. Maximum output delay results also have been obtained for proposed circuits and it has been observed that with little concession in delay, power consumption has reduced considerably.

## Authors

**Manoj Kumar** received M. Tech.degree from Guru Nanak Dev University, India in 2003. He is an Assistant Professor in the Department of Electronics & Communication Engineering Department, Guru Jambheshwar University of Science & Technology, Hisar, India. Presently he is working towards his Ph.D degree from Department of Electronics & Communication Engineering, Guru Jambheshwar University of Science & Technology, Hisar, INDIA. His research interests include low power CMOS system, Integrated circuit designs and microelectronics. He is a Life Member of IETE (India), ISTE (India) and Semiconductor Society of India.

**Dr. Sandeep K. Arya** is Associate Professor and Head in the Department of Electronics & Communication Engineering Department, Guru Jambheshwar University of Science & Technology, Hisar, India. He received M. Tech. and Ph.D degree from NIT Kurukshetra. He has more than 17 years of experience in teaching and research. His current area of research includes Optical Communication System, Integrated circuit Fabrication and CMOS circuit design. He has published more fifteen papers in referred international/national journals. He has also published more than twenty research articles in national and international conferences. He has completed project Up-gradation of Power Electronics and Industrial Control Laboratory, sponsored by MHRD, Govt. Of India at NIT Jalandhar,India.

**Dr. Sujata Pandey** received the Masters degree in electronics (VLSI) from Kurukshetra University in 1994 and the Ph.D degree from Department of Electronics, University of Delhi South Campus in Microelectronics in 1999. She joined Semiconductor Devices Research Laboratory, University of Delhi in 1996 under the Project of CSIR, Ministry of Science and Technology, Govt. of India. She joined Department of Electronics and Communication Engineering, Amity School of Engineering and Technology in 2002 as Assistant Professor. She is now Professor in the Department of Electronics and Communication Engineering, Amity University, Nodia, INDIA. She has published more than 40 research papers in International/National Journals/ Conferences. Her current research interest includes modeling and characterization of HEMTs, SOI Devices, and low power CMOS integrated circuit design. She is member of IEEE and Electron Device society.